\documentclass[twocolumn,aps,prl,showpacs,superscriptaddress]{revtex4-1}
\usepackage[latin9]{inputenc}
\usepackage{amsmath}
\usepackage{amssymb}
\usepackage{graphicx}

\makeatletter

\usepackage{color}
\usepackage{bm}
\usepackage{float}

\makeatother

\begin{document}

\title{Synergistic Longitudinal Acceleration and Transverse Oscillation in High-order Harmonic Generation}

\author{X. F. Shen}

\affiliation{Institut f\"ur Theoretische Physik I, Heinrich-Heine-Universit\"at D\"usseldorf,
40225 D\"usseldorf, Germany}

\author{A. Pukhov}
\email[Correspondence should be addressed to: ]{pukhov@tp1.uni-duesseldorf.de}


\affiliation{Institut f\"ur Theoretische Physik I, Heinrich-Heine-Universit\"at D\"usseldorf,
40225 D\"usseldorf, Germany}

\author{B. Qiao}

\affiliation{Center for Applied Physics and Technology, HEDPS, SKLNP, and School of Physics, Peking University, Beijing, 100871, China}

\date{\today}
\begin{abstract}
We propose and demonstrate that relativistic harmonics with a slowly decaying power law are generated from a femtosecond lase pulse incident parallel to a micro-scale overdense plasma. 
It is shown that due to the excitation of a strong surface wave, dense electron nanobunches are continuously accelerated forward while oscillating in the transverse laser field. Even around the stationary phase point, relativistic gamma factors of the nanobunches increase considerably, leading to a much stronger attosecond burst, compared to the case with constant gamma.   
Our two-dimensional particle-in-cell simulations and analytical theory show that this synergistic function promises a power-law harmonic spectrum $I_n/I_0 = n^{-1}$. This is much flatter than the other well-known radiation mechanisms and paves the way to unprecedentedly large energy attosecond pulses.

\end{abstract}


\maketitle
High-order harmonics generated by the interaction of intense laser pulses with overdense plasma are a promising approach to compact brilliant x-ray sources \cite{Gibbon1996,Lichters1996,Gordienko2004,Gordienko2005,Baeva2006,Dromey2006,Pukhov2006,Teubner2009,Brugge2010}, which  
have broad applications in strong field and ultrafast science \cite{Brabec2000,Krausz2009}. Compared to gas-based sources, plasma could support far higher laser intensities and offer a route to large-energy attosecond pulses. Hitherto, several radiation mechanisms of high-order harmonic generation (HHG) have been theoretically and experimentally distinguished, including coherent wake emission (CWE) \cite{Quere2006,Thaury2007}, relativistically oscillating mirror (ROM) \cite{Gibbon1996,Lichters1996,Gordienko2004,Gordienko2005,Baeva2006,Pukhov2006,Dromey2006} and coherent synchrotron emission (CSE) \cite{Brugge2010,Dromey2012,Baumann2019,Cherednychek2016,Xu2020,Edwards2020,Zhang2020}.

Among these mechanisms, CWE always dominates at weakly relativistic laser intensity ($I_0\sim10^{18}\,{\rm W/cm^2}$), the spectrum of which has a cutoff at the plasma frequency defined by the maximum electron density \cite{Thaury2007}. In strongly relativistic conditions, ROM is usually predominant, which can be understood as a periodic Doppler upshifted reflection. The harmonic spectrum could extend to much higher frequencies, following a power law with an exponent -8/3. This type of spectrum has been analytically described by the theory developed by Baeva, Gordienko and Pukhov (BGP) \cite{Baeva2006} and also observed in experiment \cite{Dromey2006}. However, due to the fast decaying power law, the intensities of harmonics in large photon energy area are rather low.

Recently, CSE as an enhanced radiation mechanism is discussed widely, which could occur when a relativistically intense laser pulse drags out a dense electron nanobunch from a sharp plasma boundary \cite{Brugge2010,Pukhov2010}.  
Here, the reflected field could far exceed the peak value of the incident one, rather than just be a simple phase modulation as in ROM. The power law of the CSE harmonic spectrum is characterized by an exponent of -4/3 to -6/5, which is much flatter than the -8/3 power from ROM. However, the CSE is highly sensitive to changes of laser and plasma parameters. Even in simulations, it does not occur in every case and a simple general description of the required conditions is still absent \cite{Brugge2010,Cherednychek2016,Xu2020}. In experiment, to observe distinct CSE spectra, a high-contrast, few-cycle laser pulse is required to keep the scale length of plasma density short enough during the whole interaction process, which is still a great challenge for present intense laser systems \cite{Dromey2012,Shen2017}. To date, no experiment has observed CSE as a dominant HHG mechanism in the reflection direction.

\begin{figure}
	\includegraphics[width=8.6cm]{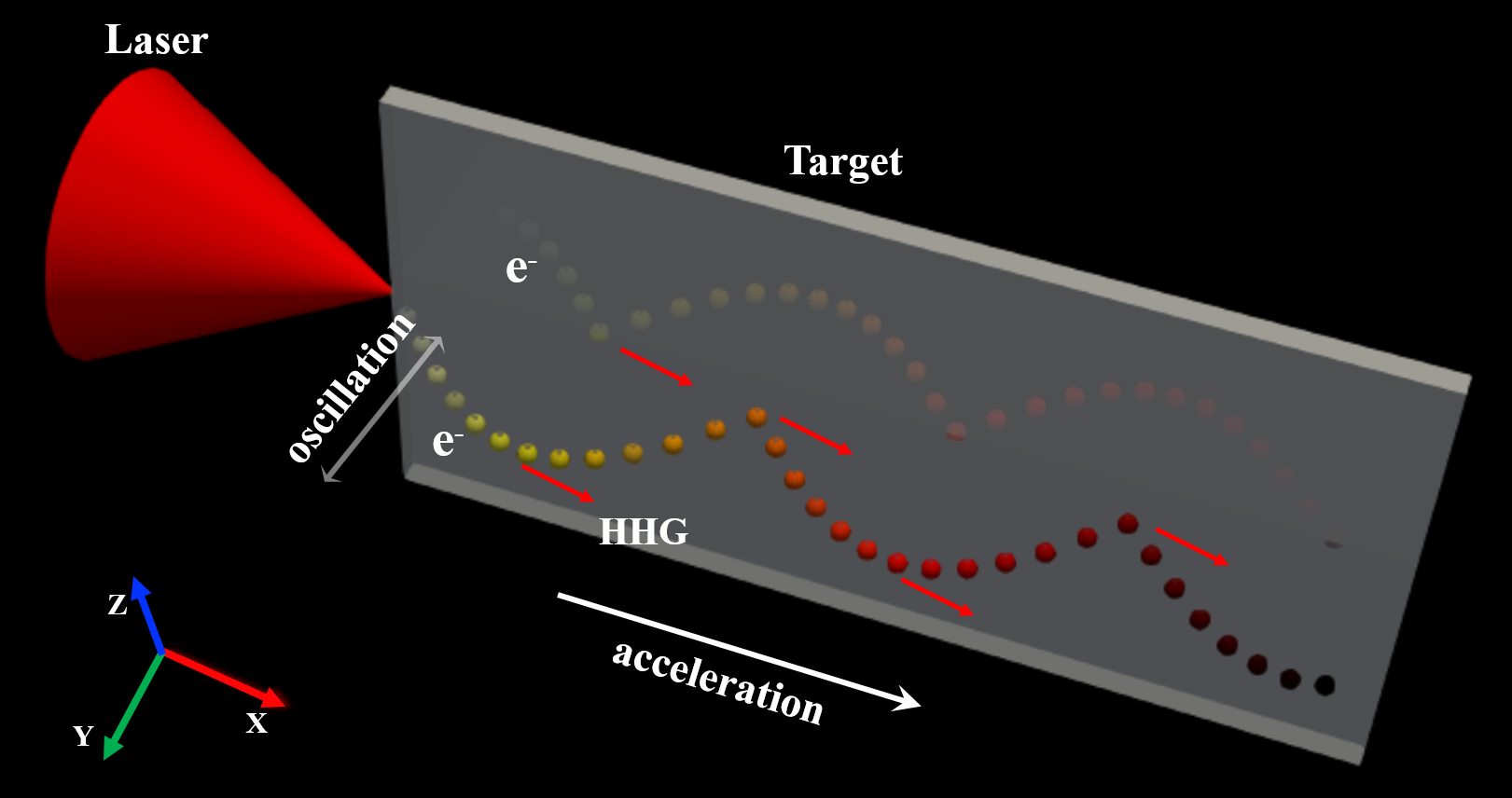} \caption{(color online) Schematic of the HHG mechanism through an intense laser pulse (red cone) irradiating a longitudinal orientated plate (grey cuboid). The trajectories of two representative electrons from both sides are shown by the colored spheres, where the color represents the relative electron energies (yellow is low and dark red is high). Intense harmonics are mainly emitted at every stationary phase (turning point), the directions of which are indicated by the red arrows. 
	}
	\label{fig:schematic} 
\end{figure}

In this Letter, we propose a highly efficient radiation mechanism through the interaction of an intense laser pulse with a tens of micrometers long overdense plasma, as shown by the schematic Fig. \ref{fig:schematic}.  
As the laser pulse propagates along the plate, dense electron nanobunches are repeatedly dragged into the vacuum by the laser field and continuously accelerated forward by the excited surface wave field \cite{Riconda2015,Pitarke2007}. 
Therefore, around the stationary phase point, relativistic gamma factors of the nanobunches still get considerable increments by the longitudinal field of surface wave, which are quite different from the scenario in the CSE mechanism.
Due to this synergistic function of longitudinal acceleration and transverse oscillation, a more slowly decaying harmonic spectra is obtained. We will refer to this mechanism as ``synergistic  synchrotron emission" (SSE) hereinafter.  
Two-dimensional (2D) particle-in-cell (PIC) simulations and analytical theory show that the power law of this radiation spectra is characterised by an exponent $-1$, which is even much flatter than that of CSE. 

The proposed regime is demonstrated using 2D PIC simulations with the Virtual Laser Plasma Lab (VLPL) code \cite{VLPL_CERN}, which applies a field solver free of numerical dispersion in the $x-$direction \cite{Pukhov2020}. A longitudinally orientated plasma plate is used, the main part of which has the uniform electron density $n_e=30n_c$, the longitudinal length $l=28\lambda$ and the transverse radius $r=0.5\lambda$. Here $n_c=m_e\omega_0^2/4\pi e^2$, where $\omega_0$ represents laser frequency, $m_e$ and $e$ are the electron mass and charge, respectively. We assume realistic laser pulses and put preplasma around the plate. Preplasma at the front has 1$\lambda$ linear density ramp and that at the lateral surfaces has linear density ramp with the $0.25\lambda$ scale. To save computational resources, a  
moving window is used. The simulation window has dimensions of $10\lambda\times20\lambda$ with $4000\times4000$ cells in $x$ and $y$ directions, respectively, which starts moving at $t=10T_0$ and stops at $t=40T_0$ ($T_0=2\pi/\omega_0$). Each cell contains 64 electrons, while the ions are assumed immobile during the short pulse duration. The $y$-polarized laser pulse with wavelength $\lambda=800$nm and peak intensity $I_0=8.65\times10^{20}\;{\rm W/cm^2}$ (corresponding to normalized amplitude $a_0=20$) is used to irradiate the plate. The transverse profile of the laser pulse is Gaussian with beam waist radius $r_L=3\lambda$. The temporal profile is cosine with a full width at half maximum (FWHM) duration $\tau_L=3T_0$ to adapt the length of the window. At the right boundary, we register fields leaving the simulation box.

\begin{figure}[b]
	\includegraphics[width=8.6cm]{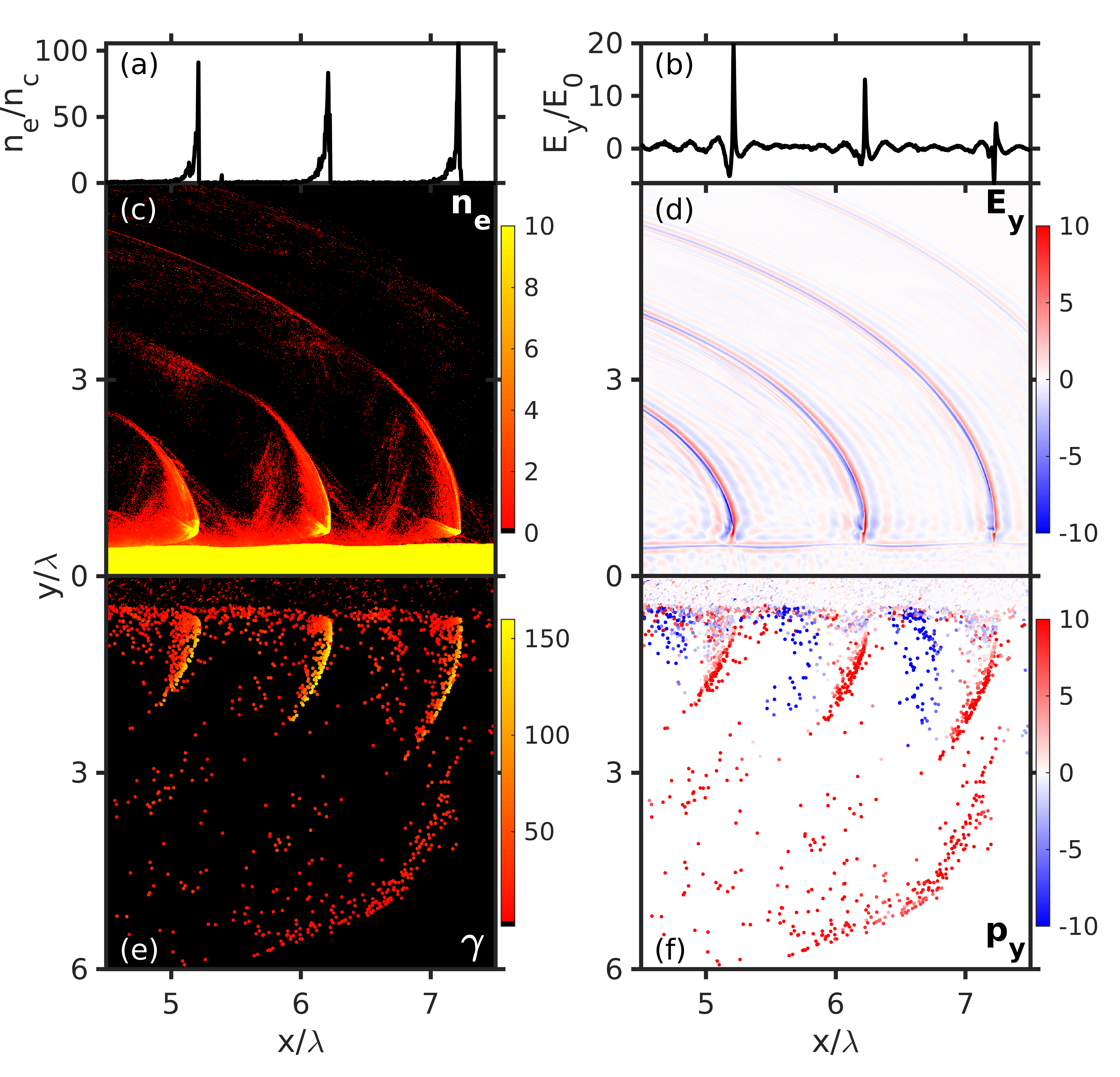} \caption{(color online) 2D PIC simulation demonstrates that dense relativistic electron nanobunches form and emit giant harmonics during a laser pulse irradiates parallel to a plate. (c)-(f) show the distributions of electron density $n_e$, attosecond pulses $E_y$ obtained from a high-pass filter, electron $\gamma_e$ and transverse momentum $p_y$ at t=$5T_0$, respectively. (a) and (b) display the electron density and radiation field at $y=0.705\lambda$, respectively.
	}
	\label{fig:fig2d} 
\end{figure}

Figure \ref{fig:fig2d} shows the simulation results at $t=5T_0$, where we select three nanobunches with relatively higher densities and energies in the upper half of simulation box to illustrate the physical processes. Here $t=0$ represents the time when the peak intensity of the laser pulse reaches the plasma. During the laser pulse interacting with the plate, significant amounts of electrons are compressed and then extracted out from the front surface of the plate, forming a series of dense electron nanobunches, separated by a laser wavelength on each side, as shown in Figs. \ref{fig:fig2d}(a) and (c) \cite{Kaymak2016,Moreau2020}. Meanwhile, a relativistic surface wave is excited, which propagates along the plate at a velocity close to the light speed $c$ and has a strong longitudinal field component \cite{Riconda2015}. Therefore, the nanobunches are continuously accelerated forward through synergistic direct laser acceleration \cite{Pukhov1999} and surface wave acceleration \cite{Riconda2015}. Fig. \ref{fig:fig2d}(c) shows the density distribution of the nanobunches and Fig. \ref{fig:fig2d}(a) corresponds to that at $y=0.705\lambda$ along $x$ direction. We can see that the electron density distribution of each nanobunch forms a highly dense ($\sim100n_c$) and very narrow $\delta$-like peak with a width of only several nanometers. The relativistic gamma factors of these dense nanobunches are high, reaching up to $\gamma_e>150$ [Fig. \ref{fig:fig2d}(e)].

\begin{figure}
	\includegraphics[width=7.6cm]{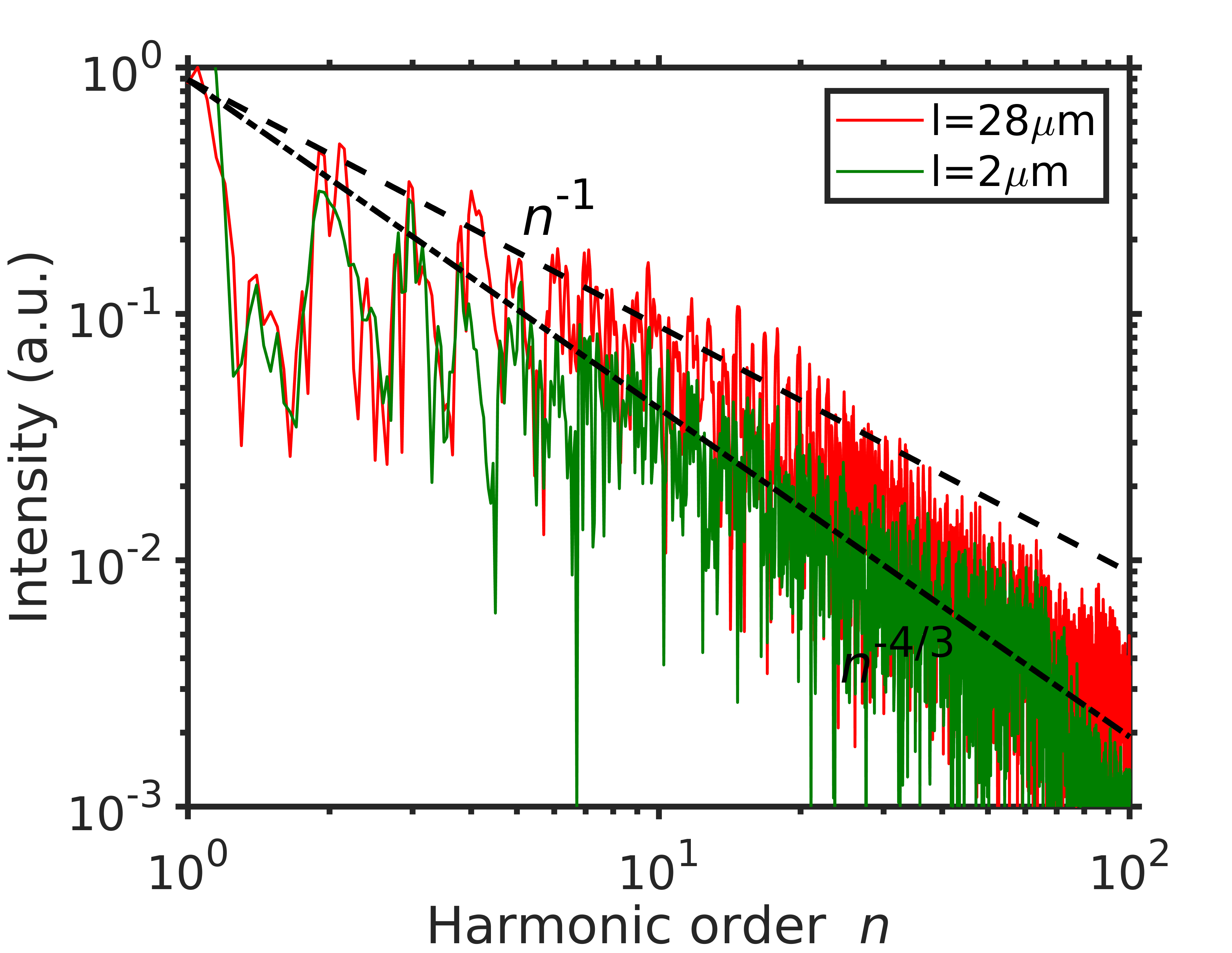} \caption{(color online) Spectrum of high-order harmonics (red line) obtained numerically for the case of $a_0=20$ and $l=28\lambda$. The dashed and dashed-dotted black lines show the predictions of the power laws with exponents $-1$ and $-4/3$, respectively. Moreover, the green line represents the harmonic spectrum obtained for the case of a short plasma plate  with $l=2\mu m$ and other parameters kept the same. Here the spectra are normalized to the same value.
	}
	\label{fig:spectra} 
\end{figure}

The distribution of electron transverse momentum $p_y$ is shown in Fig. \ref{fig:fig2d}(f). It shows that the transverse current ($j_y=en_ep_y/\gamma_e$) changes sign around the high density regions.
According to the theory developed by an der Br\"ugge and Pukhov in Ref. \cite{Brugge2010}, such dense and relativistic electron nanobunches emit synchrotron emission coherently when the transverse current approaches zero. The standard CSE harmonics spectrum is characterized by a -4/3 power law.
However, as the red line shown in Fig. \ref{fig:spectra}, the harmonics spectrum in our scheme deviates from the -4/3 
power law (dashed-dotted black line). In fact, the spectrum of SSE follows a more slowly decaying power law with an exponent close to $-1$, as shown by the dashed black line. 
This is also much flatter than the other known mechanisms. At the hundredth harmonics, the efficiency of HHG promised by SSE is about five times higher than that from CSE and three-orders of magnitude higher than that of ROM. Therefore, an extremely intense attosecond pulse burst is expected to happen. Fig. \ref{fig:fig2d}(d) shows the distribution of harmonics electric field $E_y$, where the low-frequency components below 5$\omega_0$ have been filtered out. The peak value of $E_y$ is still comparable to that of the incident laser.

Moreover, the green line in Fig. \ref{fig:spectra} represents the harmonic spectrum obtained for the case of a short plasma plate with $l=2\mu m$ and other parameters unchanged. The fields are registered at the same propagation length to exclude the effects of vacuum propagation. Comparing with the red line, it is clear that with longer plate, the intensities of high-order harmonics increase significantly, which demonstrates that this flatter spectrum should come from long scale laser-plasma interaction and the underlying physics needs to be revealed.

\begin{figure}
	\includegraphics[width=8.6cm]{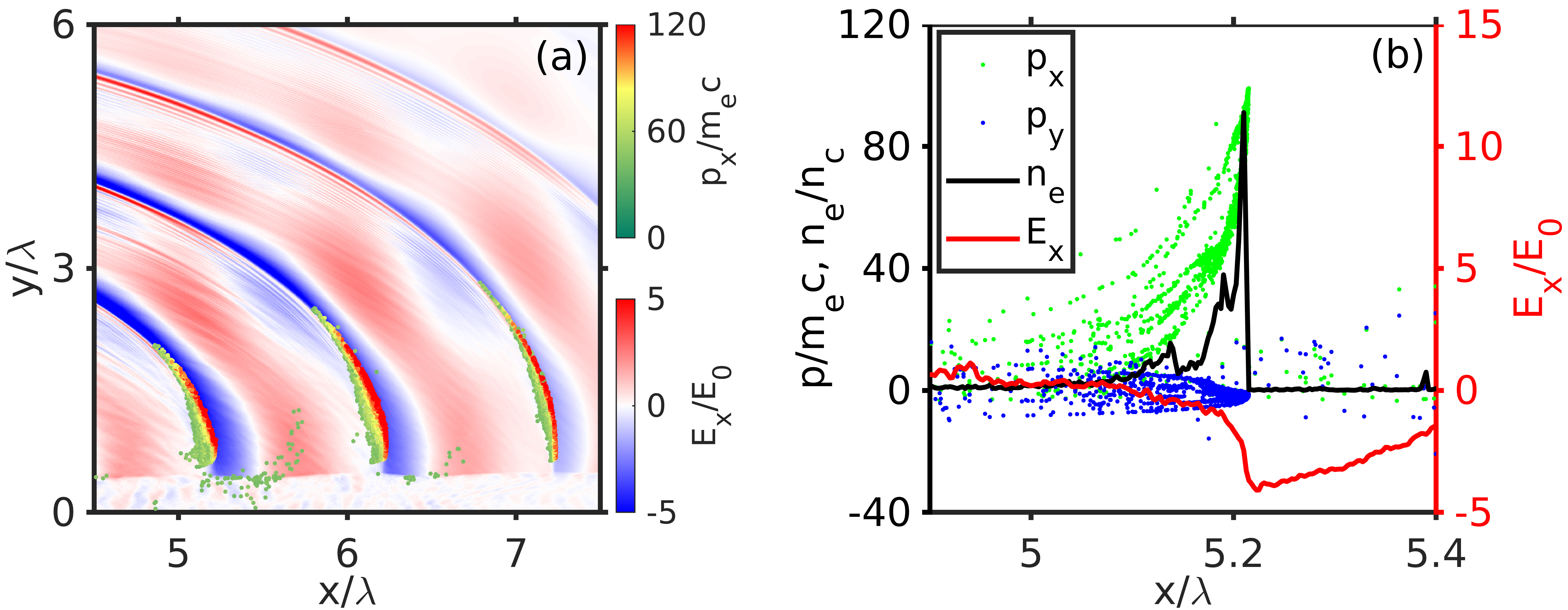} \caption{(color online) (a) the distributions of longitudinal electric field $E_x$ (blue-red) and electron momentum $p_x$ (colored dots) at $t=5T_0$. (b) the black and red lines show the distributions of electron density $n_e$ and $E_x$ at $y=0.705\lambda$, respectively. The green and blue dots represent the phase spaces of the longitudinal $p_x$ and transverse $p_y$ momenta of electrons, respectively. Note that in (a) only 1$\%$ of electrons with $\gamma_e>40$ are shown.}
	\label{fig:fig4} 
\end{figure}

In CSE, at the stationary phase point, not only the transverse current changes sign, but also the longitudinal electric field $E_x$ approaches zero, which makes the longitudinal acceleration felt by nanobunch being almost zero ($E_x+v_yB_z\simeq0$) \cite{Brugge2010,Mikhailova2012}. In other words, it is reasonable to assume $\gamma_e$ is constant there. 
Nevertheless, in our new scheme, though the electron densities of nanobunches are highly overdense, with the peak density about $100n_c$, it is still at least one order of magnitude lower than that in CSE \cite{Brugge2010,Xu2020}. Therefore, the nanobunches are not dense enough to shield the strong surface wave field completely.  
Fig. \ref{fig:fig4}(a) shows distributions of the surface wave field $E_x$ and phase space of $p_x$, where we can see that the high energy part of each nanobunch is compressed to a thin layer by the accelerating field $E_x$. Further, Fig. \ref{fig:fig4}(b) is a zoom in around the first nanobunch at $y=0.705\lambda$. It clearly shows that $E_x$ reaches its maximum of about $2\times10^{11}\,{\rm V/cm}$ close to the peak of the nanobunch. 
Such a strong $E_x$ accelerates the nanobunches forward continuously, even around the stationary phase point. 

\begin{figure}
	\includegraphics[width=8.4cm]{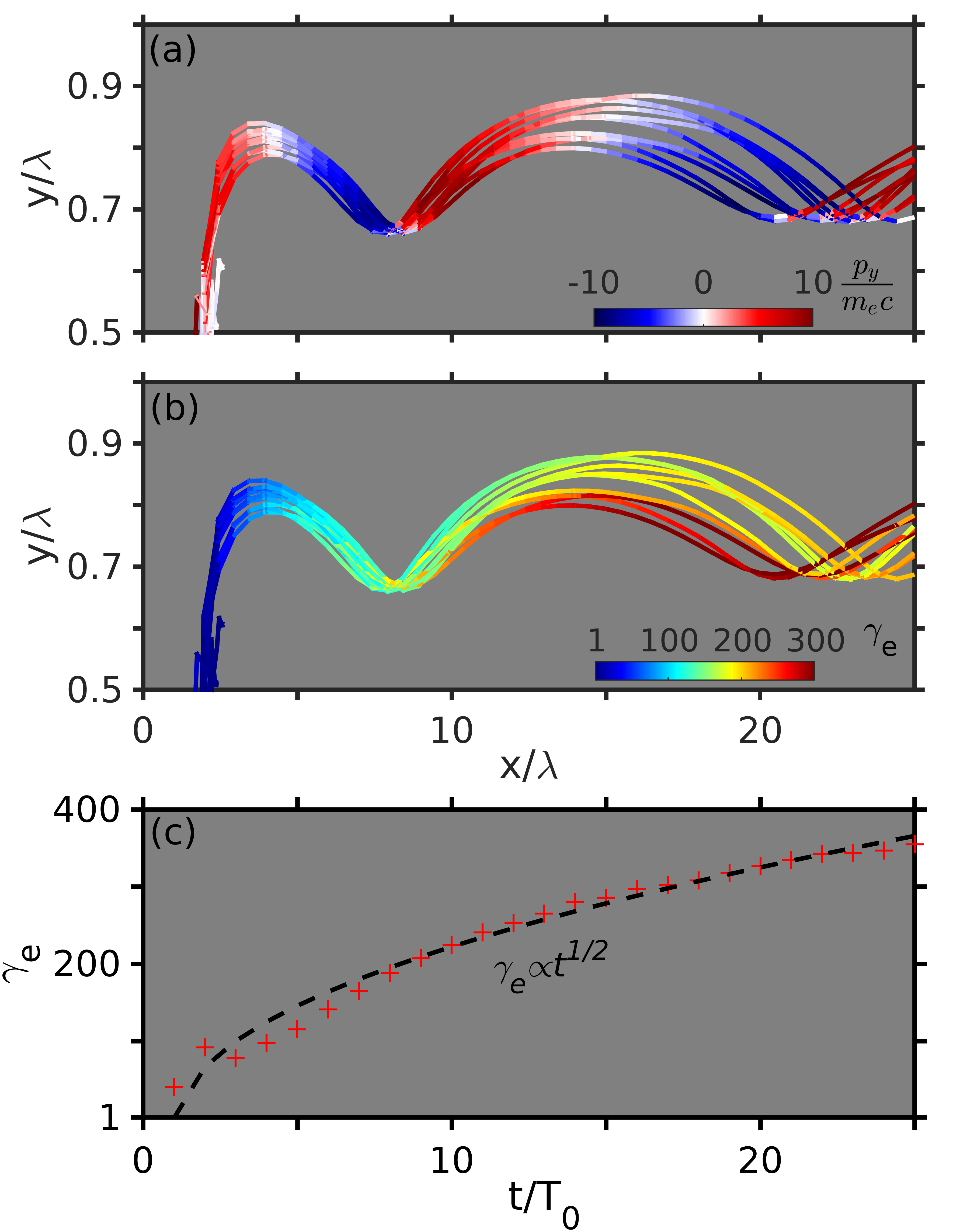} \caption{(color online) Trajectories of ten representative electrons emitting synchrotron radiation with transverse momentum $p_y$ (a) and electron $\gamma_e$ (b). The red crosses in (c) show the evolution of the maximum electron $\gamma_e$ from the first bunch of Fig. \ref{fig:fig2d}(b) and the black dashed line corresponds to the best fit.}
	\label{fig:fig5} 
\end{figure}

Figs. \ref{fig:fig5}(a) and (b) show the evolutions of $p_y$ and $\gamma_e$ of ten typical electrons, respectively. We can see that these electrons oscillate in the transverse direction while moving forward. 
Around the first ($x\sim4\lambda$) and second ($x\sim8\lambda$) turning points ($\pm 0.5T_0$), $\gamma_e$ increases about $20\%$ and $10\%$, respectively. 
Though in such ultrarelativistic regime, the increase of $\gamma_e$ barely brings any changes to the value of the absolute velocity, it still can greatly enhance the attosecond burst, since the synchrotron radiation power $P\propto\gamma_e^4$ \cite{Jackson1998}.
Meanwhile, in Fig. \ref{fig:fig5}(c), the red crosses show the evolution of the maximum $\gamma_e$ for the first nanobunch, which satisfies $\gamma_e\propto 1+t^{1/2}$ (black dashed line). Thus, the assumption of constant $\gamma_e$ is no longer valid here.

To explain the much flatter power law obtained here, we extend the theory developed by an der Br\"ugge and Pukhov to include the contribution of the time-dependent $\gamma_e$ around the stationary phase point.

First, we also start from the radiation field generated by a 1D current \cite{Brugge2010}
\begin{equation}
E_{sy}=\frac{2\pi}{c}\int j\left(t+\frac{x-x'}{c},x'\right)dx'.
\end{equation}
To include more realistic cases, we assume $j(t,x)=j(t)f[x-x_{el}(t)]$, where $f(x)$ is the function of electron distribution and $x_{el}$ is the position. After Fourier transform and considering the retarded time, we have the integral $\widetilde{E}_{sy}\left(\omega\right)=2\pi c^{-1}\widetilde{f}\left(\omega\right)\int j\left(t\right){\rm exp}\left\{-i\omega\left[t+x_{el}\left(t\right)/c\right]\right\}dt$, where $\widetilde{f}\left(\omega\right)$ is the Fourier transform of the shape function. 

Then, we need to consider this integral around the region where the phase $\Phi=\omega\left(t+x_{el}/c\right)$ is approximately stationary, i.e., $d\Phi/dt\approx0$, which constitutes the main contributions for high $\omega$. 
However, different from the assumption of a constant $\gamma_e$ in CSE, here according to the simulation results [Fig. \ref{fig:fig5}(c)], we assume $\gamma_e=1+\alpha_0\sqrt{t+t_k}$ to take the effects of the longitudinal acceleration into consideration. Note that to make it consistent with the previous theory, here  $t=0$ represents the time of the stationary phase point and $t=-t_k$ corresponds to the beginning of laser-plasma interaction or the time of the last stationary phase point. According to Fig. \ref{fig:fig2d}(f) and \ref{fig:fig5}(a), $p_y$ changes sign at the stationary phase point. Therefore, we Talor expand $p_y=\alpha_1 t$, and then get   
$x_{el}=v_0t+\alpha_2t^2$, where $\alpha_2=-\alpha_1^2/4\alpha_0^2$.  
After some algebra, the integral can be expressed as $\widetilde{E}_{sy}\left(\omega\right)\simeq C\widetilde{f}\left(\omega\right)\omega^{-1/2}$, where $C=-{i^{-1/2}}\left(\pi\alpha_2^{-1}\right)^{3/2}\alpha_3{\rm exp}\left(i\omega\alpha_2^{-1}\right)$  
is a complex prefactor, with $\alpha_3=2n_eec^{-1}\alpha_1\alpha_0^{-1}t_k^{-1/2}\left(1+1/2\alpha_2t_k\right)$. Finally, we find the spectral envelope
\begin{eqnarray}
I\left(\omega\right)&\propto&\left|\widetilde{f}\left(\omega\right)\right|^2\omega^{-1}.
\label{Eq:spectral}
\end{eqnarray} 
Here the terms $\omega^n$ with the exponent $n<-1$ are ignored, since for HHG, their contributions are relatively small. 

To compare with the simulation results, the density profile of nanobunches is assumed to be the Gaussian function $f\left(x\right)={\rm exp}[-\left(x/\delta\right)^2]$. After Fourier transformation, we have $|f\left(\omega\right)|^2={\rm exp}[-\left(\omega/\omega_{rf}\right)^2]$. Therefore, the spectral cutoff is determined mainly by $\omega_{rf}$, corresponding to the nanobunch width. Further, we choose $\omega_{rf}=80\omega_0$ ($\delta=0.0028\lambda$) to fit the PIC simulation spectrum in Fig. \ref{fig:spectra}, which matches well with the measured nanobunch width $\delta=0.0033\lambda$ [see Fig. \ref{fig:fig2d}(a)].  The dashed blue line in Fig. \ref{fig:spectra} represents the analytical spectrum from Eq. \ref{Eq:spectral}, which fits excellently with the simulation result (red line). Note that here our spectral cutoff is not directly related to $\gamma_e$, since $\gamma_e>>100$ and the electron nanobunches move in the same direction as the laser pulse. On the other hand, the width of a nanobunch is actually related to $\gamma_e$, since with larger $\gamma_e$, the longitudinal velocity is  closer to $c$, and then $\delta$ has more chance to be smaller.

In conclusion, a novel radiation mechanism of HHG has been identified from a laser pulse irradiating along an overdense plasma. Due to the synergistic longitudinal acceleration and transverse oscillation, we observed the flattest harmonic spectrum known so far, the power law of which is characterized by the exponent $-1$. According to our simulation results, SSE could tolerate preplasmas with scale lengths at a micrometer level. This should be much easier for an experimental validation, compared to the CSE mechanism. Moreover, in Fig. \ref{fig:schematic}, we just show one possible method towards the experimental realization, where an infinitely long plate along $z$ direction can be used \cite{Shen2020}. Besides this, it can also be verified through intense laser irradiating nanowires, which has been recently implemented in experiment \cite{Curtis2018}.

\section*{Acknowledgements}

This work is supported by the DFG (project PU 213/9) and GCS J\"ulich
(project QED20). X.F.S. gratefully acknowledges support by the Alexander
von Humboldt Foundation. X.F.S. acknowledges helpful discussions with
Ke Jiang, L. Reichwein, V. Kaymak and C. Baumann at HHU.\bigskip{}

\end{document}